\documentstyle [aps,prl,multicol,epsfig]{revtex}

\begin{document}

\title{Suppression of weak localization effects in low-density metallic 2D 
holes}

\author{
A. P. Mills, Jr., A. P. Ramirez, X. P. A. Gao, L. N. Pfeiffer, K. W. West 
and S. H. Simon}

\address{Bell Labs, Lucent Technologies, 600 Mountain Avenue, Murray Hill, NJ 
07974}

\maketitle

\begin{abstract}
We have measured the resistivity $R$ in a gated high-mobility GaAs two 
dimensional hole sample with densities in the range (7-17)$ \times $10$^{9}$ 
cm$^{-2}$ and at hole temperatures down to 5$ \times $10$^{-3}$E$_{F}$. We 
measure the weak localization corrections to the conductivity $g \equiv 
R^{-1}e^{2}/h$ as a function of magnetic field ($\Delta g=0.019 \pm 
$0.006 at $g$=1.5 and $T$=9 mK) and temperature ($\partial {\rm{ln}} g/\partial 
T<0.0058$ and 0.0084 at $g=1.56$ and 2.8). These values are less than a few 
percent of the value 1/$\pi $ predicted by standard weak localization theory 
for a disordered 2D Fermi liquid.
\end{abstract}

\begin{multicols}{2}
Established theory predicts that all two-dimensional systems of 
non-interacting fermions in zero magnetic field and with finite disorder 
become insulating as they approach zero temperature [1-5]. Experiments, on 
the other hand, have shown a range of 2D carrier densities for which the 
resistivity $R(T)$ decreases with decreasing temperature $T$ at the lowest 
temperatures measured [6]. At lower densities $R(T)$ increases with decreasing 
$T$ and at a particular density $R(T)$ is found to be nearly $T$ independent [7]. 
While the experiments could indicate the existence of a new metallic state 
at $T=0$ separated from the insulating state by a $T=0$ (quantum) phase 
transition, the data could also be explained within conventional theory by 
including the effects of $T$ dependent impurity scattering [8]. Exploring this 
phenomenon in a high mobility hole doped GaAs sample, we have reached 
temperatures low enough to show that the temperature independence of $R(T)$ 
extends over a wide range of densities. The difficulty of explaining, within 
a conventional theory, a temperature independent $R(T)$ over a range of 
densities leaves open the possibility of the new metallic state of matter 
suggested by the original experiments [6].

The essential idea of weak localization theory is that constructive 
interference of time reversed paths (coherent backscattering) for current 
carrying particles moving in the presence of random impurities reinforces 
the probability amplitude of the particles staying near their initial 
positions, and thus reduces the conductivity. The set of constructively 
interfering paths is confined by the rate of loss of phase coherence $\tau_\varphi^{-1}$ 
to paths of length $(D\tau_\varphi )^{1/2}$, and is 
further restricted by the phase accumulated due to the vector potential if a 
magnetic field is present. For a single band of noninteracting fermions in 
2D with a 2-fold spin degeneracy, the predicted change in electrical 
conductivity g (expressed throughout this paper in units of $e^{2}/h$), 
assuming $g \gg 1$, is [9-12], 
\begin{equation}
\Delta 
g  \approx   - (1/\pi )[\psi (\case{1}{2} + \beta /x)  -  \psi (\case{1}{2} + 
y/x)] . 
\end{equation}
Here $\psi (z)  \approx  {\rm ln}z-(2z)^{-1}-(12z^{2})^{-1}$  is the 
digamma function; $x=g^{2}H/H_{1}$; $H$ is the magnetic field; 
$H_{1}=phc/e$; and $p$ is the hole density. The momentum scattering rate is 
$\tau ^{-1} = 4\pi E_{F}/hg$, where $E_{F}=kT_{F}$ is the Fermi 
energy; and the ratio of the coherence loss and momentum scattering rates is 
$y=\tau /\tau_\varphi $. The total elastic scattering rate is $\tau 
_{0}^{-1}$, with $\beta =(\tau /\tau _{0}) \approx 1$ for short 
ranged potential scattering and $\beta >1$ otherwise. 

The largely geometrical effect which is the basis for Eq 1 makes contact 
with a real physical system when we attempt to insert the temperature 
dependence of the coherence loss rate, $\tau_\varphi^{-1}(T)$. If we assume, 
as is usually done, that $\tau_\varphi^{-1} \propto T^{q}$, Eq 1 
exhibits the well known logarithmic dependence on temperature at low values 
of the magnetic field, with $\Delta g  \approx  - (q/\pi )\rm{ln}(T)$. 
In the simplest model, in which electron-electron interactions cause 
dephasing, $\tau_\varphi =\tau_\varphi ^{ee}(T) \equiv hg/[4\pi 
^{2} kT{\rm ln}(g/2)]$, and $y=\pi {\rm{ln}}(g/2)(kT/E_{F}) \equiv T/T_\varphi $. 
For $x \ll 1$ we then have
\begin{equation}
\Delta 
g  \approx  (1/\pi )[{\rm ln}\{T/T_\varphi \} - {\rm ln}\{\beta \} + 
f_{2}(H/H_\varphi ) + \ldots ] 
\end{equation}
where $H_\varphi =H_{1} g^{-2} T/T_\varphi $ and the function 
$f_{2}(x)={\rm ln}(x)+\psi(\frac{1}{2}+1/x)=x^{2}/24+ \ldots$. Since $H_\varphi $ is a 
function of temperature, the ln$T$ dependence of $\Delta g$ in Eq 2 ceases for 
$T< T_{min} = g^{2} T_\varphi H/H_{1}$. At constant 
$H \ll H_{1} /g^{2}$ we may write [12]
$$
\nonumber
\Delta g(T)-\Delta g(T_\varphi ) \approx \frac{1}{\pi}[\psi (\case{1}{2}+ 
T/T_{min})+{\rm ln}(g^{2}H/H_{1})] 
$$

Perturbation theory shows that the effect of interactions tends to make a 2D 
system more metallic, but the perturbation expansion is only valid for 
interaction strengths characterized by $r_{s} <1$ [13], where $r_{s}$ is the 
Wigner-Seitz radius, the mean spacing between particles in units of the 
effective Bohr radius. Thus at present there are no clear predictions of 
what interaction effects there might be in our low density system for which 
$r_{s} \approx 15 \times [m^*/0.18] \times [10^{10} 
cm^{-2}/p]^{1/2} >11$, $m^*$ being the hole effective mass.

In this paper we will interpret our measurements as follows: First, the 
magnetic field dependence of the conductivity allows us to deduce $H_\varphi 
$ and hence $\tau_\varphi $, with the latter being consistent with${\tau_\varphi} =\tau_\varphi ^{ee}(T)$,
 as expected in the simple model of 
electron-electron dephasing. Second, we find that the magnetic field 
dependent weak localization correction, while definitely present, is 
suppressed by a factor of $\sim$30. Third, we find that the weak 
localization temperature dependence of the conductivity, expected based on 
the measured value of $\tau_\varphi $, is suppressed by at least the same 
amount.

Our measurements were performed on a back-gated hole-doped GaAs sample made 
from one of the wafers used in our previous study [14], a (311)A GaAs wafer 
using Al$_{x}$Ga$_{1-x}$As barriers (typical $x=0.10$) and symmetrically 
placed Si delta-doping layers above and below a pure GaAs quantum well of 
width 30 nm. The sample was thinned to $ \approx $150 $\mu $m and prepared 
in the form of a Hall bar, of approximate dimensions (2.5$ \times $9) 
mm$^{2}$, with diffused In(5\%Zn) contacts. The hole-density was varied from 
3.8 to 17$ \times $10$^{9}$ cm$^{-2}$ by means of a gate at the back of the 
sample. The zero gate bias density was 1.2$ \times $10$^{10}$ cm$^{-2}$ and 
the density at which we observe the zero magnetic field ($H = 0$) 
metal-to-insulator (MI) transition is roughly 6$ \times $10$^{9}$ cm$^{-2}$. 
The measurement current ($\sim $100 pA, 2 Hz) was applied along the 
[\underline {2}33] direction. Independent measurements of the longitudinal 
resistance per square, $R_{xx}$, from contacts on both sides of the sample 
were made simultaneously as the temperature or applied magnetic field was 
varied. The sample has a low temperature hole mobility for zero gate bias 
$\mu =2 \times 10^{5}$ cm$^{2}$V$^{-1}$s$^{-1}$. The sample was mounted 
in a top-loading dilution refrigerator.

\begin{figure}[htb]
\hspace*{20pt}\epsfig{file=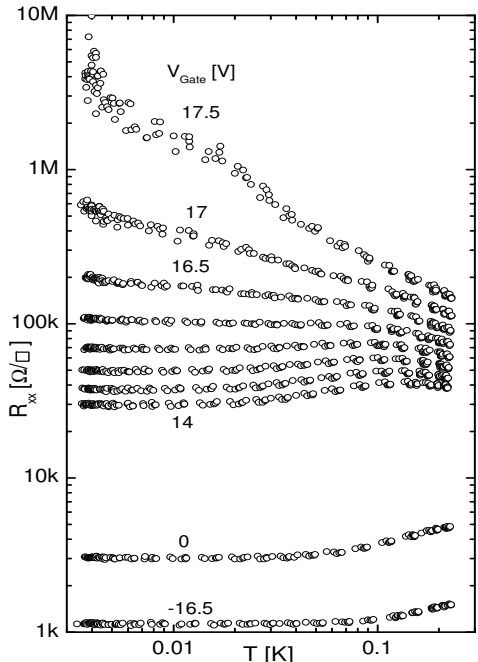,width=2.5in,height=3.5in,clip=}
\caption{Resistance per square as a function of temperature for 2D hole 
sample at various gate biases. The temperature is read from the Ge 
resistance thermometer attached to the refrigerator mixing chamber.}
\end{figure}

The most critical part of our experiment is the determination of the 
temperature $T_{H}$ of the 2D holes for a given sample lattice temperature 
$T_{L}$. The refrigerator temperature $T_{R}$ was measured by a germanium 
resistance thermometer calibrated from the 4 mK base temperature to 320 mK 
by in situ He-3 capacitance thermometry, the primary standard at such 
temperatures [15]. With commercial electronics, we were unable to obtain 
measurements on the sample at such temperatures. We used instead, spatially 
compact, well-shielded low power custom instrumentation. With this 
instrumentation we observed that at low carrier densities, for which the 
sample exhibits insulating behavior, the sample resistivity $R(T$) keeps 
increasing as $T_{R}$ is decreased to the base temperature [see Fig. 1]. 
From this we infer that $T_{L}$ is not significantly different from $T_{R}$ 
at sufficiently low measurement power levels. According to the calculation 
of Chow et al. [16], the power radiated by the 2D hole gas at temperature 
$T_{H}$ and zero lattice temperature is $P=(T^{4}/g) \times 6.75 \times 
10^{-6}$ W cm$^{-2}$ K$^{-4}$. From this we would infer that at our base 
temperature $T_{L}$= 4mK, the hole gas temperatures for $g=1$ and $g=10$ are 5.6 
and 9.4 mK respectively at our measurement power level of less than 5$ 
\times $10$^{-15}$ W cm$^{-2}$. Our measurements [17] suggest that the Chow 
et al. estimate is about a factor of two too large for our sample, but that 
there is an additional cooling rate proportional to $T^{2}$ [16]. This 
revises the estimated minimum hole temperatures to 5.0 and 7.5 mK at $g=1$ and 
10 respectively.

\begin{figure}[htb]
\hspace*{20pt}\epsfig{file=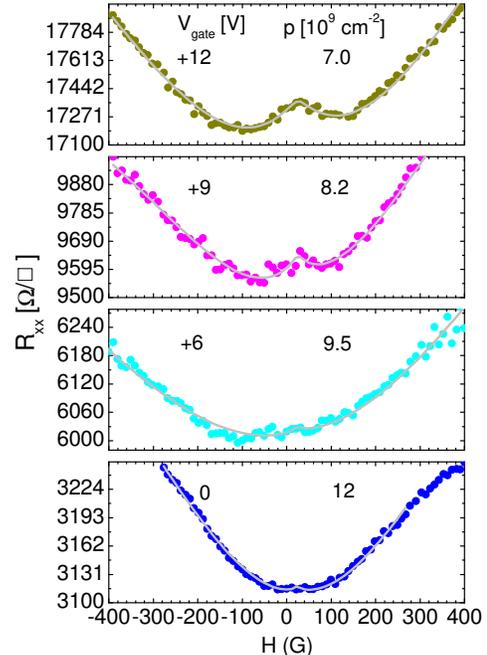,width=2.5in,height=3.5in,clip=}
\caption{Variation of longitudinal resistance with perpendicular magnetic 
field for 2D hole sample at various gate biases. The solid grey lines are 
the fit described in the text. The width of the zero field resistance peak is a 
measure of the dephasing rate.}
\end{figure}

The noise power delivered to the sample due to its being connected to the 
measuring apparatus is calculated as follows. The Johnson noise power from 
each of the six measurement leads attached to the sample is 
$P_{J}=4kT_{lead} f R_{lead}/R_{samp}=[R_{lead}/R_{\Box}] \times 6.4 
\times 10^{-14}$ W, where $R_{lead}=3.2 \Omega $ per lead, $T_{lead}{ 
\approx} $200K, $f$=8MHz and where the sample resistance is about twice the 
sample resistivity $R_{\Box}$ given in ohms per square. Normalizing to the 0.23 
cm$^{2}$ total sample area, the total Johnson noise input to the sample from 
the six leads is $P_{J}=[3k\Omega /R_{\Box}] \times 1.1 \times 10^{-15}$ 
W/cm$^{2}$. This is the only significant source of Johnson noise and raises 
the temperature of the holes in the quantum well of our sample by less than 
2 mK for all values of $R$ encountered in this study.

We determined our carrier density from Shubnikov-de Haas (SdH) 
magnetoresistance measurements obtained by averaging the longitudinal 
resistances per square $R_{xx1}$ and $R_{xx2}$ obtained from both sides of 
the sample. We plot the SdH minima vs. gate bias and fit the data using a 
straight line $P(V_{gate})=p_{0}-\alpha V_{gate}$ and the relation 
$P(V_{gate})=(2.418 \times 10^{10}$ cm$^{-2}) \times H_{\nu =1}(V_{gate}$). 
The measured slope is $\alpha $=(0.420$ \pm $0.005)$ 
\times $10$^{9}$ cm$^{-2}$V$^{-1}$ independent of sample history and the 
gate capacitance is $e \alpha $=(67.3$ \pm $0.7)pF cm$^{-2}$. The zero gate 
bias density is $p_{0}$=(12.0$ \pm $0.5) $ \times $10$^{9}$ cm$^{-2}$, where 
the variation reflects a gradual increase over the course of six months 
time.

We exhibit in Fig 2 the low field magnetoresistance of our sample at hole 
temperatures of $\sim $9 mK and for gate biases of 0, +6, +9 and +12V. The 
data were obtained over 14 sweeps of the magnetic field over the range $ \pm 
$400G and averaged into 10G bins [18]. At the lowest density $p=7 \times 
$10$^{9}$ cm$^{-2}$ there is a clear peak near zero field. We model the data 
using the function
$$
R(H)= R_{0} - R_{c} \left[\frac{(H/H_c)^2}{1+(H/H_{c})^{2}}\right] + \Delta 
 Rf_{2}(H/H_\varphi) +\eta H 
$$
shown as the solid grey lines in Fig 2.  The second term on the right hand side is a Lorentzian representing the classical  magnetoresistance [19]. The third term has the form expected from weak 
localization according to Eq 2. The last term is included to account for the 
drift associated with temporal fluctuations of $R_{0}$ [18]. The field $H$ is 
measured relative to the residual field $H_{0}$. The data for 
$V_{gate}$=+12V was fitted to obtain the following values of the free 
parameters: $R_{0}$=(17365$ \pm $14) $\Omega /\Box$; $H_{0}$=(24.8$ \pm $2.9) G; 
$R_{c}$=(4995$ \pm $1270) $\Omega/\Box $; $H_{c}$=(777$ \pm $155) G; $\eta 
$=(0.36$ \pm $0.03) ($\Omega /\Box$)G$^{-1}$; $H_\varphi $=(7.8$ \pm $3.2) G; and 
$\Delta R$=(215$ \pm $68) $\Omega/\Box $. The data in the remaining panels of 
Fig 1 were fitted with the thus determined value of $H_{0} $(24.8G) and with 
$H_\varphi $ scaled from $H_\varphi $=7.8G as $H_\varphi  \propto p/g^{2}$. The 
fitted values of $\Delta g$ are consistent with a constant $\Delta g$=0.01, 
much less than the weak localization prediction $\Delta g=\pi ^{-1}$. 
The expected value of $H_\varphi =H_{1} g^{-2} T/T_\varphi $ using $T_\varphi 
=(E_{F}/k)/(\pi $ln$\{g/2\})$ is not well defined for the conditions of 
Fig 1a where $g=1.50$. We use the results of Ref 16 to approximate 
$g/$ln$\{g/2\} \approx $4.16 and obtain $H_\varphi $=12 G. This result implies 
that the shape of the weak localization magnetoresistance correction is 
approximately as expected for an ordinary 2D metal, although the magnitude 
of the weak localization correction is more than an order of magnitude 
smaller than expected.

The fact that the fitted value of $H_\varphi $ agrees roughly with the 
expectation that $T_\varphi =(E_{F}/k)/(\pi $ln$\{g/2\})$, implies that the 
simple model of electron-electron interaction causing dephasing $\tau_ \varphi 
=\tau_\varphi ^{ee}(T)$ 
is roughly correct. We may then also 
compute the temperature $T_{min} = g^{2} T_\varphi  H/H_{1}= 
[1$mK$][H/8.436 G][0.18/m^*][g^2/{\rm ln}\{g/2\}]$ below which the weak localization 
correction to the resistivity stops being proportional to ln$T$. For $H<$2G, 
$T_{min} <$16 mK at $g$=10 and $<6$ mK at $g$=5. This means that we should be able 
to see a weak localization contribution to the conductivity given the 
precision of our knowledge of the magnetic field zero from Fig 2.

\begin{figure}[htb]
\hspace*{20pt}\epsfig{file=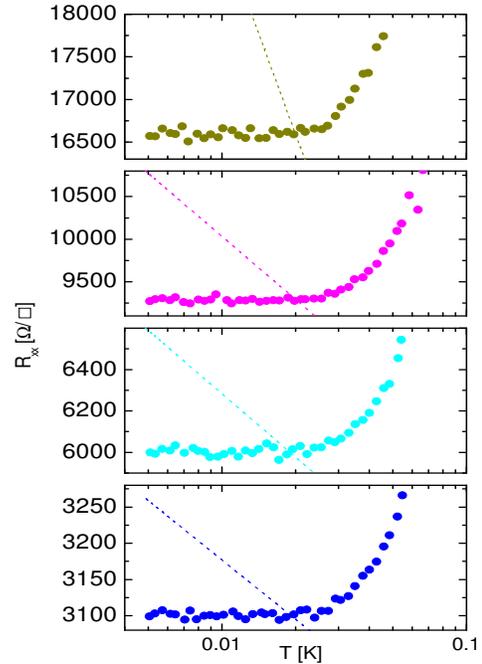,width=2.5in,height=3.5in,clip=}
\caption{Temperature dependence of the longitudinal resistivity of a 2D hole 
gas for various gate biases. The dashed lines are approximate weak localization 
predictions according to Eq 2.}
\end{figure}

Fig 3 shows a measurement of $R(T)$ over the range 5 to 100 mK for gate biases 
of 0, +6, +9, and +12V. The data were obtained with the 25G residual field 
of the superconducting solenoid cancelled to within the $ \pm $2.5 G 
precision of our fitted value of $H_{0}$. The excitation levels were in all 
cases 5 fW/cm$^{2}$ or less, thus ensuring that the minimum hole temperature 
is less than 6 mK for the highest resistivity curve. Likewise, the minimum 
hole temperature is less than 8 mK for the 3 k$\Omega $/ measurement. The 
dashed lines, calculated by simply adding $(1/\pi)\log(T/20{\rm mK})$ to the low temperature conductivity $g(20 {\rm mK})$, are clearly 
at variance with the measurements. Rather, at 
temperatures well below the characteristic temperature 0.2$T_{F}$ associated 
with the activated resistance bump, the resistance becomes nearly constant. 
Fitting an expression $g(T)=e^2/(h R(T_{0}))+b $ln$\{T/T_{0}$\}to 
the data for $T<T_0 \equiv20$ mK, we find 99\% confidence upper limits of $b<0.0058$, 
0.0084, 0.018, 0.015, and 0.033 for densities $p=$ 7.0, 8.2, 9.5, 12, and 
14.5$ \times $10$^{9}$ cm$^{-2}$, and corresponding $R(T_{0})$=16500, 
9290, 6000, 3100, and 1950 $\Omega $. These upper limits for $b$ are one to 
nearly two orders of magnitude smaller than the $b=1/\pi$ expected according 
to Eq 2.

At this time there is no satisfactory explanation for a suppression of weak 
localization effects in a 2D electron gas. Weak localization effects in $R(T)$ 
of the expected magnitude have been observed in a Si MOSFET sample at 
electron densities of order 3$ \times $10$^{12}$ cm$^{-2}$ by Pudalov et al. 
[20], and in a GaAs sample at hole densities of order 10$^{11}$ cm$^{-2}$ by 
Hamilton et al. [21]. This suggests that interaction effects associated with 
the large value of r$_{s}$ in our experiment may be playing an important 
role [13]. 

The first author is grateful to C. M. Varma for encouraging him to pursue 
these experiments and for useful conversations. The authors benefited 
greatly from the assistance of D. S. Greywall in calibrating the Ge 
resistance temperature scale using He-3 melting curve thermometry.

\end{multicols}


\begin{references}
\bibitem{17} Further verification that the holes were not heated by the current used 
to measure the resistance was obtained from studying the apparent nonlinear 
resistance of our sample at $H=0$. We measured $R_{T}(T)$[see Fig 1], the 
longitudinal resistance-per-square $R_{xx}$ as a function of the temperature 
$T_{Ge}$ (deduced from the Ge thermometer) and gate bias and $R_{P}(P)$, the 
resistance as a function of the excitation power per unit area P at the 
minimum temperature of the sample, $T_{min}{\approx} $ 4 mK. We then 
construct the function $T_{eff}(P)=R_{T}^{-1} (R_{P}(P))$ which gives 
the effective temperature $T_{eff}$ corresponding to a power $P$. If we assume 
that our measured $T_{eff}(P)$ is consistent with the non-linear resistance 
being caused by $I^{2}R $ heating of the hole carriers and 
that $T_{eff}$ is the same as the hole temperature, we conclude from our 
measurements that, for base temperature sample resistivities of 30, 100 and 
600 $k\Omega /\Box$, the hole temperatures are below 5 mK when the sample 
lattice temperature is 4 mK and we use excitation levels of less than 5 
fW/cm$^{2}$. When the base temperature resistivity is 3 $k\Omega /\Box$, we were 
unable to directly determine the cooling rate for $T<30$ mK. Extrapolation of 
the fitted segment to low power leads us to conclude that, for a base 
temperature sample resistivity of 3 $k\Omega / \Box$, the hole temperature is 
below 8 mK when the sample lattice temperature is 4 mK and we use excitation 
levels of less than 5 fW/cm$^{2}$.
\bibitem{18} At positive gate bias, the resistance fluctuates at fixed gate bias and 
field with a wide range of characteristic times. The variations are 
presumably manifestations of universal conductance fluctuations [see Ref. 3] 
associated with the motion of impurities.
\end{references}
\end{document}